\documentclass{jpsj3}
\usepackage{txfonts}

\def\gsim{\mathop {\vtop {\ialign {##\crcr 
$\hfil \displaystyle {>}\hfil $\crcr \noalign {\kern1pt \nointerlineskip } 
$\,\sim$ \crcr \noalign {\kern1pt}}}}\limits}
\def\lsim{\mathop {\vtop {\ialign {##\crcr 
$\hfil \displaystyle {<}\hfil $\crcr \noalign {\kern1pt \nointerlineskip } 
$\,\,\sim$ \crcr \noalign {\kern1pt}}}}\limits}

\title{$T/B$ Scaling in $\beta$-YbAlB$_4$}

\author{\name{Shinji \name{Watanabe}}$^1$ and \name{Kazumasa \name{Miyake}}$^2$}
\inst{$^1$Department of Basic Sciences, Kyushu Institute of Technology, Kitakyushu, Fukuoka 804-8550, Japan \\
$^2$Toyota Physical and Chemical Research Institute, Nagakute, Aichi 480-1192, Japan \\} 

\abst{
The scaling behavior over four decades of the ratio of temperature $T$ to magnetic field $B$ 
observed in the magnetization in $\beta$-YbAlB$_4$ is theoretically examined. 
By developing a theoretical framework that exhibits the quantum critical phenomena of Yb-valence 
fluctuations under a magnetic field, 
we show that the $T/B$-scaling behavior can appear near the quantum critical point of the valence transition. 
The emergence of the $T/B$ scaling indicates the presence of the small characteristic energy scale 
of critical Yb-valence fluctuations. 
It is argued that 
the quantum valence criticality offers a unified explanation for the 
unconventional quantum criticality  
as well as the $T/B$ scaling in $\beta$-YbAlB$_4$. 
}

\begin{document}
\maketitle

Quantum critical phenomena in itinerant electron systems 
that do not follow the conventional spin-fluctuation theory~\cite{Moriya,MT,Hertz,Millis}
have attracted attention in condensed matter physics~\cite{MW2014}. 
The heavy-electron metal $\beta$-YbAlB$_4$ has recently attracted great interest
since the unconventional quantum criticality, such as 
the magnetic susceptibility $\chi\sim T^{-0.5}$, 
the electronic specific-heat coefficient $C_{\rm e}/T\sim-\log T$, 
and approximately $T$-linear resistivity,  
has been observed at low temperatures 
at least below 3~K down to a few hundred mK~\cite{Nakatsuji,Matsumoto}. 

Interestingly, from the magnetization data for $T\lsim 3$~K and 
the magnetic field $B\lsim 2$~T, 
in $\beta$-YbAlB$_4$ it has been discovered that 
the magnetic susceptibility $\chi$ shows the following $T/B$ scaling behavior over four decades 
of $T/B$:   
\begin{eqnarray}
\chi^{-1}=(\mu_{\rm B}B)^{1/2}
\varphi
\left(
\frac{k_{\rm B}T}{\mu_{\rm B}B}
\right), 
\label{eq:kaiB}
\end{eqnarray}
where $\mu_{\rm B}$ and $k_{\rm B}$ are the Bohr magneton and 
Boltzmann constant, respectively, and $\varphi$ is the function 
$\varphi(x)=\Lambda(\Gamma^2+x^2)^{1/4}$ with $\Lambda$ and $\Gamma$ being  
constants~\cite{Matsumoto}. 
Namely, $\chi^{-1}/(\mu_{\rm B}B)^{1/2}$ is expressed 
as a single scaling function of the ratio $T/B$. 

This striking behavior of Eq.~(\ref{eq:kaiB}) 
calls for theoretical explanation, and 
it has so far been proposed that 
anisotropic hybridization between f and conduction electrons is the key origin 
of the emergence of $T/B$ scaling~\cite{RCNT2012}. 
However, 
this theory requires an assumption that the renormalized f level 
is pinned at the hybridized band edge, 
and it also seems unclear whether the unconventional criticality 
observed in $C_{\rm e}/T$ and the resistivity can be explained 
by the anisotropic hybridization. 

Recently, it has been shown theoretically that 
a new type of quantum criticality emerges near the quantum critical point (QCP) 
of the first-order valence transition 
in Yb- and Ce-based heavy-electron systems~\cite{WM2010}. 
Critical valence fluctuations of Yb or Ce cause the quantum criticality 
in physical quantities such as $\chi$, $C_{\rm e}/T$, resistivity, and the NMR/NQR 
relaxation rate $(T_{1}T)^{-1}$, which give a unified explanation for the measured 
unconventional criticality in $\beta$-YbAlB$_4$~\cite{WM2010,WM2012}. 
Hence, it is interesting to examine whether the critical Yb-valence fluctuation 
can account for the $T/B$ scaling observed in $\beta$-YbAlB$_4$.  

In this Letter, we show that the $T/B$ scaling can be understood 
from the viewpoint of the quantum valence criticality. 
By developing a theoretical framework for the quantum critical phenomena of Yb-valence fluctuations 
under a magnetic field, 
we show that the $T/B$ scaling emerges near the QCP of the valence transition. 
We demonstrate that the emergence of the $T/B$ scaling is a hallmark of the presence of
the small characteristic energy scale of the critical Yb-valence fluctuations.

We employ the theoretical framework developed in Ref.~\citen{WM2010}, 
whose formulation is extended so as to describe the effect of a magnetic field. 
Hereafter, we take the energy units of $k_{\rm B}=1$, $\hbar=1$, and $\mu_{\rm B}=1$ 
unless otherwise noted.
We consider the simplest minimal model  
\begin{eqnarray}
H=H_{\rm PAM}+H_{U_{\rm fc}}+H_{\rm Zeeman}
\label{eq:Hamil}
\end{eqnarray}
as the starting Hamiltonian, 
where 
$H_{\rm PAM}=\sum_{{\bf k}\sigma}\varepsilon_{\bf k}
c_{{\bf k}\sigma}^{\dagger}c_{{\bf k}\sigma}
+\varepsilon_{\rm f}\sum_{i\sigma}n_{i\sigma}^{\rm f}
+\sum_{{\bf k}\sigma}\left(V_{\bf k}
f_{{\bf k}\sigma}^{\dagger}c_{{\bf k}\sigma}
+{\rm h.c.}
\right)
+U\sum_{i}n_{i\uparrow}^{\rm f}n_{i\downarrow}^{\rm f}
$, 
$
H_{U_{\rm fc}}=\sum_{i\sigma\sigma'}n_{i\sigma}^{\rm f}n_{i\sigma'}^{\rm c}
$, 
and the Zeeman term 
$
H_{\rm Zeeman}=-h\sum_{i}
S_{i}^{{\rm f}z}
$
with $n_{i\sigma}^{a}\equiv a_{i\sigma}^{\dagger}a_{i\sigma}$ 
for $a=f$ or $c$ 
and $S_{i}^{{\rm f}z}\equiv \frac{1}{2}(n_{i\uparrow}^{\rm f}-n_{i\downarrow}^{\rm f})$ 
in the standard notation. 

To discuss the quantum critical phenomena of Yb- (and Ce-) valence fluctuations, 
first we take into account the local correlation effect by the $U$ term, 
which is the strongest interaction in Eq.~(\ref{eq:Hamil}) 
responsible for the realization of the heavy-electron state. 
Then perturbative expansion with respect to the $U_{\rm fc}$ term is performed. 
To perform the procedure, we employ the slave-boson large-$N$ expansion method~\cite{OM2000}. 
Here we set the orbital degeneracy $N=2$ to discuss $\beta$-YbAlB$_4$, 
where the Kramers-doublet ground state is realized. 
Hence, $\sigma=\uparrow, \downarrow$ in Eq.~(\ref{eq:Hamil}) 
should be regarded as the effective ``spin" index that specifies the Kramers doublet. 
The slave-boson operator $b_{i}$ is introduced to eliminate the doubly occupied 
state for $U\to\infty$ under the constraint
$\sum_{\sigma}n_{i\sigma}^{\rm f}+2b_{i}^{\dagger}b_{i}=1$. 
The Lagrangian is written as 
${\cal L}={\cal L}_{0}+{\cal L}'$, 
where ${\cal L}_{0}$ is the Lagrangian for $H_{\rm PAM}+H_{\rm Zeeman}$ 
with the term 
$-\sum_{i}\lambda_{i}\left(\sum_{\sigma}n_{i\sigma}^{\rm f}+2b_{i}^{\dagger}b_{i}-1\right)$ 
with $\lambda_{i}$ being the Lagrange multiplier 
and ${\cal L}'$ is the Lagrangian for $H_{U_{\rm fc}}$ (see Ref.~\citen{WM2010} for detail). 

For $\exp(-S_{0})$ with the action $S_{0}=\int_{0}^{\beta}d\tau{\cal L}_{0}(\tau)$, 
the saddle-point solution is obtained via the stationary condition $\delta{S}_{0}=0$ 
by approximating spatially uniform and time-independent solutions, i.e., 
$\lambda_{\bf q}=\lambda\delta_{\bf q}$ and $b_{\bf q}=b\delta_{\bf q}$. 
The solution is obtained by solving the mean-field equations 
$\partial{S}_{0}/\partial\lambda=0$ and $\partial{S}_{0}/\partial{b}=0$ self-consistently. 

\begin{figure}
\includegraphics[width=6cm]{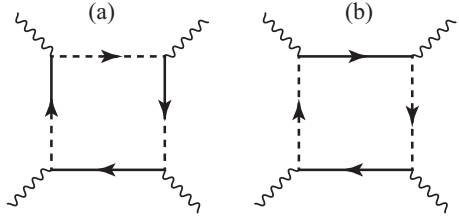}
\caption{Feynman diagrams for the (a) first term and (b) second term 
in the mode-coupling constant $v_{4\sigma}$ given by Eq.~(\ref{eq:v4}). 
The solid and dashed lines with an arrow represent f and conduction-electron 
Green functions $G^{\rm ff}_{{\bf k}\sigma}({i}\varepsilon_{n})$ and 
$G^{\rm cc}_{{\bf k}\sigma}({i}\varepsilon_{n})$, respectively. 
The half-dashed and solid line with an arrow represents the off-diagonal Green function
$G^{\rm cf}_{{\bf k}\sigma}({i}\varepsilon_{n})$. 
The wiggly line represents critical valence fluctuations.
}
\label{fig:v4}
\end{figure}

For $\exp(-S')$ with the action $S'=\int_{0}^{\beta}d\tau{\cal L}'(\tau)$, 
we introduce the identity applied by a Stratonovich-Hubbard transformation 
$
e^{-S'}
=\int{\cal D}\varphi\exp\left[
\sum_{i\sigma}\int_{0}^{\beta}d\tau\left\{
-\frac{U_{\rm fc}}{2}\varphi_{i\sigma}(\tau)^2
+{i}\frac{U_{\rm fc}}{\sqrt{2}}
(c_{i\sigma}f_{i\sigma}^{\dagger}-f_{i\sigma}c_{i\sigma}^{\dagger})
\varphi_{i\sigma}(\tau)
\right\}
\right]
$.
The partition function is expressed as 
$
Z=\int{\cal D}\varphi\exp(-S[\varphi]) 
$
with $S=S_{0}+S'$. 
By performing Grassmann number integrations for $cc^{\dagger}$ 
and $ff^{\dagger}$, we obtain 
$
Z=\int{\cal D}\varphi\exp(-S[\varphi])
$
with
\begin{eqnarray}
S[\varphi]=\sum_{\sigma}
\left[
\frac{1}{2}\sum_{\bar{q}}\Omega_{2\sigma}(\bar{q})
\varphi_{\sigma}(\bar{q})\varphi_{\sigma}(-\bar{q})
+\sum_{\bar{q}_{1},\bar{q}_{2},\bar{q}_{3}}
\Omega_{3\sigma}(\bar{q}_{1},\bar{q}_{2},\bar{q}_{3})
\right. 
& &
\nonumber
\\
\left.
\times
\varphi_{\sigma}(\bar{q}_{1})\varphi_{\sigma}(\bar{q}_{2})\varphi_{\sigma}(\bar{q}_{3})
\delta\left(
\sum_{i=1}^{3}\bar{q}_{i}
\right)
+\sum_{\bar{q}_{1},\bar{q}_{2},\bar{q}_{3},\bar{q}_{4}}
\Omega_{4\sigma}(\bar{q}_{1},\bar{q}_{2},\bar{q}_{3},\bar{q}_{4})
\right.
& &
\nonumber
\\
\left.
\times
\varphi_{\sigma}(\bar{q}_{1})\varphi_{\sigma}(\bar{q}_{2})
\varphi_{\sigma}(\bar{q}_{3})\varphi_{\sigma}(\bar{q}_{4})
\delta\left(
\sum_{i=1}^{4}\bar{q}_{i}
\right)
+\cdots
\right], 
\label{eq:S}
& &
\end{eqnarray} 
where the abbreviation 
$\bar{q}\equiv({\bf q},{i}\omega_{l})$ with 
$\omega_{l}=2l\pi{T}$ is used. 
Since the long wavelength around ${\bf q}={\bf 0}$ and 
the low-frequency regions play dominant roles in the critical phenomena, 
$\Omega_{i\sigma}$ for $i=2,3$, and 4 are expanded for $q$ and $\omega_{l}$ 
around $({\bf 0},0)$:
$
\Omega_{2\sigma}(q,{i}\omega_{l})\approx
\eta_{0\sigma}+A_{\sigma}q^2+C_{\sigma}\frac{|\omega_{l}|}{q}
$, 
where
$
\eta_{0\sigma}=U_{\rm fc}\left[
1-U_{\rm fc}
\left\{
\chi_{0\sigma}^{\rm ffcc}({\bf 0},0)
-\chi_{0\sigma}^{\rm cfcf}({\bf 0},0)
\right\}
\right].
$
Here  
$
\chi_{0\sigma}^{\alpha\beta\gamma\delta}({\bf q},{i}\omega_{l})
=-\frac{T}{N_{\rm s}}\sum_{{\bf k},n}
G_{{\bf k}+{\bf q}\sigma}^{\alpha\beta}({i}\varepsilon_{n}+{i}\omega_{l})
G_{{\bf k}\sigma}^{\gamma\delta}({i}\varepsilon_{n}), 
$
where 
$G_{{\bf k}\sigma}^{\rm ff}({i}\varepsilon_{n})=1/[{i}\varepsilon_{n}
-\bar{\varepsilon}_{{\rm f}\sigma}-\bar{V}_{\bf k}^{2}/({i}\varepsilon_{n}-\bar{\varepsilon}_{{\bf k}\sigma})]$, 
$G_{{\bf k}\sigma}^{\rm cc}({i}\varepsilon_{n})=1/[{i}\varepsilon_{n}
-\bar{\varepsilon}_{{\bf k}\sigma}-\bar{V}_{\bf k}^{2}/({i}\varepsilon_{n}-\bar{\varepsilon}_{{\rm f}\sigma})]$, 
and 
$G_{{\bf k}\sigma}^{\rm cf}({i}\varepsilon_{n})=\bar{V}_{\bf k}/[({i}\varepsilon_{n}-\bar{\varepsilon}_{{\rm f}\sigma})({i}\varepsilon_{n}-\bar{\varepsilon}_{{\bf k}\sigma})-\bar{V}_{\bf k}^{2}]$ 
with $\varepsilon_{n}=(2n+1)\pi{T}$. 
Here, $\bar{\varepsilon}_{{\bf k}\sigma}$, $\bar{\varepsilon}_{{\rm f}\sigma}$, 
and $\bar{V}_{\bf k}$ are defined as 
$\bar{\varepsilon}_{{\bf k}\sigma}\equiv\varepsilon_{\bf k}+\frac{U_{\rm fc}}{2}$, 
$\bar{\varepsilon}_{{\rm f}\sigma}\equiv\varepsilon_{\rm f}+\frac{U_{\rm fc}}{2}+\frac{\lambda}{\sqrt{N_{\rm s}}}-P(\sigma)\frac{h}{2}$, 
and 
$\bar{V}_{\bf k}\equiv\frac{V_{\bf k}b}{\sqrt{N_{\rm s}}}$, respectively,  
with $P(\uparrow)\equiv +1$ and $P(\downarrow)\equiv -1$. 
Since $\chi_{0\sigma}^{\rm ffcc}({\bf 0},0)\gg
\chi_{0\sigma}^{\rm cfcf}({\bf 0},0)$, 
as shown in Ref.~\citen{WM2010}, hereafter 
we use the approximated form 
$
\eta_{0\sigma}\approx U_{\rm fc}\left[
1-U_{\rm fc}
\chi_{0\sigma}^{\rm ffcc}({\bf 0},0)
\right]
$
for simplicity of calculation. 
For $\Omega_{3\sigma}$ and $\Omega_{4\sigma}$, expansion up to 
the zeroth order is performed as   
$\Omega_{3\sigma}(\bar{q}_1,\bar{q}_2,\bar{q}_3)
\approx v_{3\sigma}/\sqrt{\beta{N}_{\rm s}}$ 
and 
$\Omega_{4\sigma}(\bar{q}_1,\bar{q}_2,\bar{q}_3,\bar{q}_4)
\approx v_{4\sigma}/(\beta{N}_{\rm s})$, respectively. 
The mode-coupling constant $v_{4\sigma}$ is derived as  
\begin{eqnarray}
v_{4\sigma}=\frac{U_{\rm fc}^4}{4}
\left[
\frac{T}{N_{\rm s}}\sum_{n}\sum_{\bf k}G_{{\bf k}\sigma}^{\rm cf}({i}\varepsilon_{n})^2
G_{{\bf k}\sigma}^{\rm cc}({i}\varepsilon_{n})
G_{{\bf k}\sigma}^{\rm ff}({i}\varepsilon_{n})
\right.
\nonumber
\\
+\left.\frac{T}{2N_{\rm s}}\sum_{n}\sum_{\bf k}
G_{{\bf k}\sigma}^{\rm cc}({i}\varepsilon_{n})^2
G_{{\bf k}\sigma}^{\rm ff}({i}\varepsilon_{n})^2
\right], 
\label{eq:v4}
\end{eqnarray}
where the first and second terms are expressed by a Feynman diagram in Figs.~\ref{fig:v4}(a) and 
\ref{fig:v4}(b), respectively.

Since renormalization-group analysis has shown that higher order terms $v_{i} \ (i\ge 3)$ are 
irrelevant for the $d=3$ spatial dimension~\cite{WM2010}, 
we construct the action for the Gaussian fixed point. 
Taking account of the mode-coupling effects up to the 4th order in $S[\varphi]$ 
in Eq.~(\ref{eq:S}), we employ Feynman's inequality for the free energy~\cite{Feynman}:
$F\le F_{\rm eff}+T\langle S-S_{\rm eff}\rangle_{\rm eff}\equiv\tilde{F}(\eta)$, 
where 
$S_{\rm eff}$ is the effective action for the best Gaussian, 
$S_{\rm eff}[\varphi]=\frac{1}{2}\sum_{\sigma}\sum_{{\bf q},l}
\chi_{{\rm v}\sigma}({\bf q},{i}\omega_{l})^{-1}
|\varphi_{\sigma}({\bf q},{i}\omega_{l})|^2$. 
Here, $\chi_{{\rm v}\sigma}({\bf q},{i}\omega_{l})$ is the valence susceptibility 
defined as 
\begin{eqnarray}
\chi_{{\rm v}\sigma}({\bf q},{i}\omega_{l})^{-1}
\approx\eta+A_{\sigma}q^2+C_{\sigma}\frac{|\omega_{l}|}{q}, 
\label{eq:chi_v}
\end{eqnarray}
where the notation follows in Ref.~\citen{WM2010}.
Under the optimal condition $\frac{d\tilde{F}(\eta)}{d\eta}=0$,  
the self-consistent renormalization (SCR) equation under a magnetic field 
in the $A_{\sigma}q_{\rm B}^2\lsim\eta$ regime
is obtained: 
\begin{eqnarray}
& &\sum_{\sigma}A_{\sigma}q_{{\rm B}\sigma}^4\frac{T_{0\sigma}}{T_{{\rm A}\sigma}^2}
\left(
1+\frac{v_{4\sigma}q_{{\rm B}\sigma}^3}{\pi^2}\frac{T_{0\sigma}}{T_{{\rm A}\sigma}^2}
\right)
\nonumber
\\
&{\times}&
\left[
y_{0\sigma}-\tilde{y}_{\sigma}+\frac{3}{2}y_{1\sigma}t_{\sigma}
\left\{
\frac{x_{\rm c}^3}{6\tilde{y}_{\sigma}}
-\frac{1}{2\tilde{y}_{\sigma}}
\int_{0}^{x_{\rm c}}\frac{x^3}{x+\frac{t_{\sigma}}{6\tilde{y}_{\sigma}}}dx
\right\}
\right]
\nonumber
\\
&{\times}&
\left[
C_{2\sigma}+\frac{x_{\rm c}^3}{3}\frac{t_{\sigma}}{\tilde{y}_{\sigma}^2}
\int_{0}^{x_{\rm c}}\frac{x^4}{\left(
x+\frac{t_{\sigma}}{6\tilde{y}_{\sigma}}
\right)^2
}dx
\right]
=0, 
\label{eq:VSCReq}
\end{eqnarray}
where $\tilde{y}_{\sigma}=y\frac{A}{A_{\sigma}}\left(\frac{q_{\rm B}}{q_{{\rm B}\sigma}}\right)^2$, 
$t_{\sigma}=\frac{T}{T_{0\sigma}}$, 
$T_{0\sigma}=\frac{A_{\sigma}q_{{\rm B}\sigma}^3}{2{\pi}C_{\sigma}}$, 
and  
$T_{{\rm A}\sigma}=\frac{Aq_{{\rm B}\sigma}^2}{2}$ 
with $q_{{\rm B}\sigma}$ being the Brillouin zone for ``spin" $\sigma$. 
Note that $A$, $C$, and $q_{\rm B}$ are the zero-field values of $A_{\sigma}$, 
$C_{\sigma}$, and $q_{{\rm B}\sigma}$, respectively. 
Here, $y$ is defined as $y\equiv\frac{\eta}{Aq_{\rm B}^2}$, and the 
dimensionless integral variable and its cutoff are defined as    
$x\equiv q/q_{\rm B}$ and $x_{\rm c}\equiv q_{\rm c}/q_{\rm B}$, respectively.  
The parameters $y_{0\sigma}$ and $y_{1\sigma}$ are given by
\begin{eqnarray}
y_{0\sigma}&=&\frac{\frac{\eta_{0\sigma}}{A_{\sigma}q_{{\rm B}\sigma}^2}
+v_{4\sigma}\frac{T_{0\sigma}}{T_{{\rm A}\sigma}^2}\frac{q_{{\rm B}\sigma}^3}{\pi^2}C_{1\sigma}}
{1+v_{4\sigma}\frac{T_{0\sigma}}{T_{{\rm A}\sigma}^2}\frac{q_{{\rm B}\sigma}^3}{\pi^2}C_{2\sigma}},
\label{eq:y0}
\\
y_{1\sigma}&=&
\frac{v_{4\sigma}\frac{T_{0\sigma}}{T_{{\rm A}\sigma}^2}\frac{4q_{{\rm B}\sigma}^3}{3\pi^2}}
{1+v_{4\sigma}\frac{T_{0\sigma}}{T_{{\rm A}\sigma}^2}\frac{q_{{\rm B}\sigma}^3}{\pi^2}C_{2\sigma}}, 
\label{eq:y1}
\end{eqnarray}
respectively, where $C_{1\sigma}$ and $C_{2\sigma}$ are constants given by 
$C_{1\sigma}=\int_{0}^{x_{\rm c}}dxx^3\ln\left|
\frac{(A_{\sigma}q_{{\rm B}\sigma}^2x^3)^2+(C_{\sigma}\omega_{\rm c}/q_{{\rm B}\sigma})^2}{(A_{\sigma}q_{{\rm B}\sigma}^2x^3)^2}
\right|
$ 
and 
$C_{2\sigma}=2(C_{\sigma}\omega_{\rm c})^2
\int_{0}^{x_{\rm c}}dx\frac{x}{(A_{\sigma}q_{{\rm B}\sigma}^3x^3)^2+(C_{\sigma}\omega_{\rm c})^2}
$, respectively. 

Note that in the zero-field case, $h=0$, Eq.~(\ref{eq:VSCReq}) is reduced to 
the simple form
\begin{eqnarray}
y=y_{0}+\frac{3}{2}y_{1}t
\left\{
\frac{x_{\rm c}^3}{6y}
-\frac{1}{2y}
\int_{0}^{x_{\rm c}}\frac{x^3}{x+\frac{t}{6y}}dx
\right\}
\label{eq:VSCReq2}
\end{eqnarray}
with 
$y_{0}=y_{0\sigma}$, $y_{1}=y_{1\sigma}$, and $t=T/T_{0}$, where 
$T_{0}=\frac{Aq_{\rm B}^3}{2\pi{C}}$, 
which reproduces Eq.~(\ref{eq:VSCReq}) in Ref.~\citen{WM2010}. 
It is noted that 
at the QCP of the valence transition, 
the magnetic susceptibility diverges, whose singularity is the same as 
the valence susceptibility $\chi\propto \chi_{\rm v}({\bf 0},0)
\propto y^{-1}$ since the main contribution to $\chi$ and $\chi_{\rm v}$ 
comes from the common many-body effects caused by $U_{\rm fc}$, 
which can be expressed by the common Feynman diagrams near the QCP~\cite{WM2010}.  

In this Letter, we demonstrate that the $T/B$ scaling behavior appears when 
the characteristic temperature of critical valence fluctuations $T_{0}$ is 
smaller than (or at least comparable to) the measured lowest temperature. 
Hence, we here set the coefficient $A_{\sigma}$ in Eq.~(\ref{eq:chi_v}) 
as a small input parameter 
to discuss the effect of a small $T_{0}$ on physical quantities. 
The procedure of our calculation is summarized as follows. 
First, we solve the saddle-point solution for $\exp(-S_{0})$ at $T=0$ 
for given parameters of $\varepsilon_{\rm f}$, $V_{\bf k}$, $U=\infty$, and $h$ 
at the filling 
$n\equiv\frac{1}{2N_{\rm s}}\sum_{i\sigma}\langle n_{i\sigma}^{\rm f}+ 
n_{i\sigma}^{\rm c}\rangle$ 
by using the slave-boson mean-field theory. 
Second, we calculate $\chi_{0\sigma}^{\rm ffcc}({\bf 0},0)$ 
and the $\left[\dots\right]$ part in Eq.~(\ref{eq:v4}) 
by using the saddle-point solution.  
Then we obtain $\eta_{0\sigma}$ and $v_{4\sigma}$ for a given $U_{\rm fc}$. 
Third, 
by using $y_{0\sigma}$ and $y_{1\sigma}$ 
obtained from Eqs.~(\ref{eq:y0}) and (\ref{eq:y1}), respectively, 
we solve the valence SCR equation~[Eq.~(\ref{eq:VSCReq})] and finally obtain $y(t)$.

We note that 
the crystalline electronic field (CEF) ground state of $\beta$-YbAlB$_4$ 
has been suggested to be the Kramers doublet, 
which is well separated from the excited CEF levels~\cite{Nakatsuji,NC2009}. 
Since the analysis of the CEF-level scheme, which well reproduces the anisotropy of 
the magnetic susceptibility, deduces that a hybridization node exists along the $c$-axis 
in $\beta$-YbAlB$_4$~\cite{NC2009,IM1996,Aoki1998}, 
we employ the anisotropic hybridization in the form of $V_{\bf k}=V(1-\hat{k}_{z}^{2})$ 
with $\hat{\bf k}\equiv{\bf k}/|{\bf k}|$ 
to simulate $\beta$-YbAlB$_4$ most simply. 

For evaluation of the saddle-point solution, we employ the typical parameter set for heavy-electron systems: $D=1$, $V=0.65$, and $U=\infty$ at the filling $n=0.8$. Here, $D$ is the half bandwidth of conduction electrons given by $\varepsilon_{\bf k}=k^2/(2m_{0})-D$, which is taken as the energy unit. The mass $m_{0}$ is set such that the integration from $-D$ to $D$ of the density of states of conduction electrons per ``spin" and site is equal to $1$. 

Following the argument in Ref.~\citen{WM2010}, we discuss the general property at the QCP of the valence transition by defining it as the point with the solution of Eq.~(\ref{eq:VSCReq}) $y$ being zero at $T=0$, which is identified to be $(\varepsilon_{\rm f}, U_{\rm fc})=(-0.7,0.700328652)$ for $A=5\times10^{-6}$ at $h=0$. 
This $U_{\rm fc}$ is larger than $U_{\rm fc}^{\rm RPA}\equiv 1/\chi_{0}^{\rm ffcc}({\bf 0},0)=0.62404$ for $\varepsilon_{\rm f}=-0.7$, which reflects the mode-coupling effect of critical valence fluctuations. Namely, a positive $v_{4\sigma}$ overcomes a negative $\eta_{0\sigma}$ for $U_{\rm fc}>U_{\rm fc}^{\rm RPA}$ 
[see Eq.~(\ref{eq:y0})], giving rise to 
$0.700328652>U_{\rm fc}^{\rm RPA}$. 

It is noted that here we set a rather large c-f hybridization strength $V$ to simulate $\beta$-YbAlB$_4$ with a large  fundamental  characteristic energy scale $\approx 200$~K~\cite{Nakatsuji}. Actually, the characteristic energy for heavy electrons, which is defined as the Kondo temperature $T_{\rm K}\equiv \bar{\varepsilon}_{\rm f}-\mu$ within the saddle-point solution for $\exp(-S_{0})$, is estimated to be  $T_{\rm K}=0.02437$. 

To examine the magnetic-field dependence of $y(t)$ at the QCP, we solve 
the valence SCR equation [Eq.~(\ref{eq:VSCReq})] for $U_{\rm fc}=0.700328652$. 
To make a comparison with 
experiments where a magnetic field from the order of $B=10^{-4}$~T to $B=2$~T 
is applied, 
we apply a magnetic field ranging from $h=10^{-8}$ to $h=10^{-4}$~\cite{h_Lifshitz}. 
Here we note that the energy unit of our theory is the conduction bandwidth $D=1$, 
which is of the order of $10^{4}~{\rm T} \ (\approx 10^{4}~{\rm K})$. 
To compare with experiments measured in the temperature range from the order of $T=10^{-2}$~K 
to $T=3$~K, we solve the valence SCR equation [Eq.~(\ref{eq:VSCReq})] 
for $6\times 10^{-6}\le T \le 3\times 10^{-4}$. 
As noted above, $A$ is set as 
$A=5\times10^{-6}$, which gives $T_{0}=3\times10^{-6}$, 
slightly smaller than the lowest temperature but of the same order. 
Owing to the smallness of $A_{\sigma}$, hereafter we neglect its field dependence and 
set $A=A_{\sigma}$ for $h\ne 0$.  

\begin{figure}
\includegraphics[width=7.5cm]{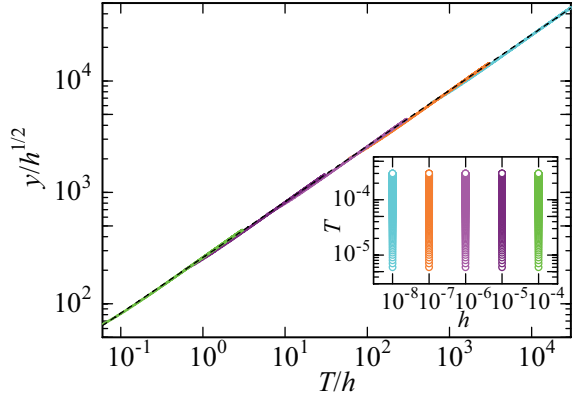}
\caption{(Color online) 
Scaling of the data for $T\le 3\times 10^{-4}$ and $h\le 10^{-4}$. 
The inset shows the $T$-$h$ range where the scaling applies.
The dashed line represents the fitting function $a(T/h)^{1/2}$. 
The data were obtained by solving the valence SCR equation [Eq.~(\ref{eq:VSCReq})] 
for $D=1$, $V=0.65$, $\varepsilon_{\rm f}=-0.7$, $U=\infty$,  
and $U_{\rm fc}=0.700328652$ at $n=0.8$. 
}
\label{fig:log}
\end{figure}

The results are shown in Fig.~\ref{fig:log}.
Intriguingly, 
we find that all the data 
over four decades of the magnetic field fall down to a single  scaling function of the ratio $T/h$: 
\begin{eqnarray}
y=h^{1/2}
\varphi\left(\frac{T}{h}\right).
\label{eq:scaling}
\end{eqnarray}
The least-squares fit of the scaling function $\varphi(x)=ax^{1/2}$ to the data for $10^1\le T/h\le 10^4$ shows that the data are well fitted by the dashed line in Fig.~\ref{fig:log}. Namely, $y/h^{1/2}\approx a(T/h)^{1/2}$, i.e., $y\approx T^{1/2}$. 
This implies that the quantum criticality of Yb-valence fluctuations is dominant, giving rise to the non-Fermi liquid regime~\cite{WM2010}. 
This behavior coincides with the measured scaling function Eq.~(\ref{eq:kaiB}) for  $x=T/h\gg \Gamma$.  
It is noted that  
as $x$ decreases the data tend to deviate from $ax^{1/2}$, i.e., there is a tendency of upward deviation from the dashed line toward $x=T/h\ll 1$ in the smaller $T/h$ region than that shown in Fig.~\ref{fig:log}. 
This reflects the suppression of the valence susceptibility by applying the magnetic field. Namely, as $x=T/h$ decreases, the crossover from the non-Fermi-liquid regime with the quantum valence criticality to the Fermi liquid regime with suppressed valence fluctuation occurs. 
As noted above, the uniform magnetic susceptibility $\chi$ has the same temperature dependence as the valence susceptibility $\chi\propto\chi_{\rm v}({\bf 0},0)\propto y^{-1}$~\cite{WM2010}. 
This indicates that general tendency of the $T/B$ scaling observed in the magnetization data of $\beta$-YbAlB$_4$ can be reproduced by the solutions of the valence SCR equation [Eq.~(\ref{eq:VSCReq})] under a magnetic field.

To analyze how the $T/h$ scaling behavior appears in the present theoretical framework, 
let us rewrite the valence SCR equation [Eq.~(\ref{eq:VSCReq})] with the scaled form of 
$y/t^{1/2}$ and $t/h$:
\begin{eqnarray}
& &\sum_{\sigma}A_{\sigma}q_{{\rm B}\sigma}^4\frac{T_{0\sigma}}{T_{{\rm A}\sigma}^2}
\left(
1+\frac{v_{4\sigma}q_{{\rm B}\sigma}^3}{\pi^2}\frac{T_{0\sigma}}{T_{{\rm A}\sigma}^2}
\right)
\left[
\left(\frac{y_{0\sigma}}{h^{1/2}}\right)
-\left(\frac{\tilde{y}_{\sigma}}{h^{1/2}}\right)
\right.
\nonumber
\\
& &
\left.
+\frac{3}{2}y_{1\sigma}\left(\frac{t_{\sigma}}{h}\right)
\left\{
\frac{x_{\rm c}^3}{6\left(\frac{\tilde{y}_{\sigma}}{h^{1/2}}\right)}
-\frac{1}{2\left(\frac{\tilde{y}_{\sigma}}{h^{1/2}}\right)}
\int_{0}^{x_{\rm c}}\frac{x^3}
{x+h^{1/2}\frac{\left(\frac{t_{\sigma}}{h}\right)}
{6\left(\frac{\tilde{y}_{\sigma}}{h^{1/2}}\right)}}dx
\right\}
\right]
\nonumber
\\
&{\times}&
\left[
C_{2\sigma}+\frac{x_{\rm c}^3}{3}
\frac{\left(\frac{t_{\sigma}}{h}\right)}{\left(\frac{\tilde{y}_{\sigma}}{h^{1/2}}\right)^2}
\int_{0}^{x_{\rm c}}\frac{x^4}{\left\{
x+h^{1/2}\frac{\left(\frac{t_{\sigma}}{h}\right)}{6\left(\frac{\tilde{y}_{\sigma}}{h^{1/2}}\right)}
\right\}^2
}dx
\right]
=0.  
\label{eq:VSCReq2}
\end{eqnarray}
We see that most terms can be expressed in the form of $y/h^{1/2}$ and $t/h$,  
except for the term with the $x$-integration in each square bracket $[\ ]$. 
Namely, extra $h^{1/2}$ factors appear in the denominators of the integrands. 
This implies that {\it the $T/h$ scaling does not hold exactly}. 
From Eq.~(\ref{eq:VSCReq2}), however, it turns out  
that in the case of $t_{\sigma}/\tilde{y}_{\sigma}\gg 1$, 
the denominators of the integrands become large, 
which make the $x$-integrations negligibly small. 
We confirmed that this is the case  
when $T_{0}$ is below (or at least comparable to) the measured lowest temperature.  
In the present calculation, we set $T_{0}=3.0\times 10^{-6}$ and 
the lowest temperature for the data in Fig.~\ref{fig:log} is $T=6.0\times 10^{-6}$, 
i.e., $T_{0}$ is a few times smaller than the lowest temperature. 
Note that $T_{0}$ is the same order as the lowest temperature. 

From these results, 
the $T/B$ scaling observed in $\beta$-YbAlB$_4$ suggests that 
a small characteristic temperature of critical valence fluctuations $T_{0}$ exists. 
Since the measured lowest temperature is on the order of $10^{-2}$~K in $\beta$-YbAlB$_4$, 
$T_{0}$ is considered to be of the same order or smaller. 

As shown in Ref.~\citen{WM2010}, 
because of the strong local-correlation effect by $U\gg D$, 
an almost dispersionless critical valence-fluctuation mode appears, 
giving rise to the extremely small $q^2$-coefficient $A$ in the momentum space.   
This almost flat mode is reflected in the emergence of the extremely small characteristic 
temperature $T_{0}$.   
Owing to the extremely small $T_{0}$, the temperature at the low-$T$ measurement 
can be regarded as a ``high" temperature in the scaled temperature $t=T/T_{0}\gsim 1$, 
where unconventional quantum criticality 
emerges in physical quantities such as $\chi$, $(T_{1}T)^{-1}$, $C_{\rm e}/T$, and 
resistivity, which well account for the behavior of $\beta$-YbAlB$_4$~\cite{WM2010}.
Our results show that observation of the $T/B$ scaling indicates the presence of 
the small characteristic temperature $T_{0}$. 
In other words, quantum valence criticality gives a unified explanation for the 
unconventional criticality in physical quantities as well as 
the $T/B$ scaling in $\beta$-YbAlB$_4$. 

To verify the existence of such a small $T_0$ experimentally, 
the measurement of the dynamical valence susceptibility $\chi_{\rm v}({\bf q},\omega)$ 
is desirable as a direct observation. 
M\"{o}ssbauer measurement and ESR measurement also seem to be possible probes 
to detect $T_0$~\cite{Kobayashi,nakatsuji2}, which are interesting future studies.

Although Eq.~(\ref{eq:kaiB}) shows that $\chi\approx (\mu_{\rm B}B)^{-1/2}$ for $x=\frac{k_{\rm B}T}{\mu_{\rm B}B}\ll\Gamma$, it should be noted that a very narrow range of experimental data is used to derive this limiting behavior: A large magnetic field of $B=2$~T and intermediate temperatures of $0.2~{\rm K}\le T \le 0.5~{\rm K}$  (but {\it not} the lowest temperature) are used~\cite{Matsumoto}.  Namely, the scaling form in the $x\ll\Gamma$ regime is an outcome of the transient behavior of the magnetization, where $\chi$ is greatly suppressed to be almost constant around $B=1-2$~T~\cite{Matsumoto}. Furthermore, we should be careful about the fact that the whole scaling range of $10^{-1}\le T/B\le10^{3}$ is {\it not} covered by a series of experimental data as a function of $T$ for a fixed $B$. From these circumstances, it seems to be appropriate to consider that $\chi^{-1}/(\mu_{\rm B}B)^{1/2}\approx(\frac{k_{\rm B}T}{\mu_{\rm B}B})^{1/2}$ for the $x=\frac{k_{\rm B}T}{\mu_{\rm B}B}\gg\Gamma$ regime in Eq.~(\ref{eq:kaiB}), derived from the experimental data for the wide $T$ and $B$ range is a substantial scaling function. 

Theoretically, as shown in Ref.~\citen{WM2008}, the location of the QCP in the ground-state phase diagram in the $\varepsilon_{\rm f}$-$U_{\rm fc}$ plane is moved by applying $h$. If the system is located in the vicinity of the QCP at $h=0$, applying $h$ moves the system away from the QCP, which causes the marked suppression of $\chi$ at large $h$. In this Letter, we discussed the $h$-dependence of $\chi$ through the $h$-dependence of $\eta_{0\sigma}$ and $v_{4\sigma}$ with the QCP being unmoved for simplicity of analysis. Taking account of this effect is expected to make the crossover $T/h$ between the Fermi liquid and non-Fermi liquid regimes shift to the larger-$T/h$ direction in Fig.~\ref{fig:log}, which is an interesting future study for quantitative analysis. 
  
We note that in the present theory
{\it the key origin of the emergence of the $T/B$ scaling is not the anisotropic hybridization but the quantum valence criticality}.
In the present calculation, the renormalized f level is not located at the band edge 
as expected in the general (and natural) situation for heavy-electron state. 
Namely, in our framework, even without the pinning of the f-level, i.e., 
the fine tuning of the f-level position, the $T/B$ scaling behavior can emerge, which is in sharp contrast to Ref.~\citen{RCNT2012}. 

We also note that the $T/B$ scaling does not hold exactly 
as discussed below Eq.~(\ref{eq:VSCReq2}). 
When $T_{0}$ is comparable to the middle-$T$ range applied to the scaling plot of the data, 
the deviation from the single scaling function shown in Fig.~\ref{fig:log} becomes visible. 
As shown in Ref.~\citen{WM2010}, the valence susceptibility, i.e., the magnetic susceptibility 
behaves as $y^{-1}\sim t^{-1/2}$ for $t\gsim1$ and $y^{-1}\sim t^{-2/3}$ for $T_{\rm K}/T_{0}>t\gg 1$.
At sufficiently high temperatures, $T\gg T_{\rm K}$, the Curie-Weiss behavior $y^{-1}\sim t^{-1}$ appears. 
Hence, we stress that the emergence of the $T/B$ scaling is an approximate outcome for the intermediate-temperature region which satisfies $t_{\sigma}/\tilde{y}_{\sigma}\gg 1$ 
as explained above.

In summary, we have shown that the $T/B$ scaling 
together with the unconventional quantum criticality observed in $\beta$-YbAlB$_4$ 
can be understood from the viewpoint of the quantum valence criticality 
in a unified way.

\begin{acknowledgment}
\acknowledgement
We acknowledge S.~Nakatsuji, Y.~Matsumoto, K.~Kuga, and H.~Kobayashi for showing us their experimental data with enlightening discussions on their analyses. 
This work was supported by Grants-in-Aid for Scientific Research (No. 24540378 and No. 25400369) from Japan Society for the Promotion of Science (JSPS). 
One of us (S.W.) was supported by JASRI (Proposal 
No. 0046 in 2012B, 2013A, 2013B, and 2014A). 
\end{acknowledgment}


\end{document}